# ANISOTROPIC BEHAVIOUR LAW FOR SHEETS USED IN STAMPING - COMPARATIVE STUDY OF STEEL AND ALUMINIUM

# LOI DE COMPORTEMENT ANISOTROPE POUR TOLES UTILISEES EN EMBOUTISSAGE - ETUDE COMPARATIVE DE L'ACIER ET L'ALUMINIUM

Jean-Jacques SINOU[+], Bruno MACQUAIRE[*]

MECANIQUE, Mécanique des solides et des structures/ Mechanics of solids and structures (M6).

[+]Laboratoire de Tribologie et Dynamique des Systèmes UMR CNRS 5513,
Ecole Centrale de Lyon, 36 avenue Guy de Collongue, 69134 Ecully, FRANCE.

[*]RENAULT S.A,
Manager BIW Advanced Design & new Materials,
Technocentre Renault,
1 avenue du golf, 78288 Guyancourt Cedex, FRANCE.

**Abstract** - For a car manufacturer, unweighting vehicles is obvious. Replacing steel by aluminium moves towards that goal. Unfortunately, aluminium's stamping numerical simulation results are not yet as reliable as those of steel. Punch-strength and spring-back phenomena are not correctly described. This study on aluminium validates the behaviour law Hill 48 quadratic yield criterion with both isotropic and kinematic hardening. It is based on the yield surface and on associated experimental tests (uniaxial test, plane tensile test, plane compression and tensile shearing).
*Solids and Structures / stamping of sheets / yield surface / mixed hardening .*

**Résumé** - Réduire le poids des véhicules est l'un des principaux objectifs des constructeurs automobiles. Le remplacement de l'acier par l'aluminium va dans ce sens. Malheureusement, la simulation numérique de l'emboutissage des tôles d'aluminium n'a pas encore atteint le niveau de fiabilité de l'acier : des problèmes tels que l'effet poinçon ou le retour élastique n'y sont pas encore correctement décrits. L'étude consiste donc à valider la loi de comportement du critère quadratique HILL 48 avec écrouissage mixte (écrouissage isotrope et cinématique) à partir de la surface de charge et des essais expérimentaux classiques (essais de traction simple, traction plane, compression plane et cisaillement).
*Solides et structures / Emboutissage des tôles / surface de charge / écrouissage mixte.*

*Version française abrégée*

Actuellement, l'un des principaux objectifs des constructeurs automobiles est l'allégement des véhicules. Or la tendance actuelle ne va pas dans ce sens : équipement standard de plus en plus complet, confort optimisé, assistance électronique et renforts de structure omniprésents. Cela contribue à une augmentation du poids des véhicules. Se trouvant, de plus, confronté aux règlements imposant une réduction de la consommation des véhicules, le secteur automobile se doit d'axer ses recherches sur l'allégement des structures. L'aluminium se positionne donc aujourd'hui comme un concurrent potentiel de l'acier : on peut espérer un gain de poids de 30-40% par rapport à l'acier. La diversité des matériaux implique que chacun d'entre eux soit employé de façon optimale. Ainsi, la mise en forme des matériaux métalliques de l'automobile passe par la maîtrise de l'emboutissage. En effet, parmi les procédés de mise en forme, l'emboutissage des tôles minces est l'un des plus répandus dans l'industrie automobile. L'enjeu des prochaines années est de concevoir et valider des outils d'emboutissage grâce à la simulation numérique, sans avoir recours aux essais de validation actuels qui nécessitent la construction d'outils d'essai coûteux. Cependant, la simulation numérique ne peut valider des calculs sur la faisabilité de pièces de carrosserie qu'à la condition que les paramètres caractérisant les matériaux soient les plus justes possible et reflètent au mieux la réalité. C'est pour cette raison que les codes de calcul nécessitent des modèles de comportement réalistes et des données fiables sur les matériaux.

L'objectif de ce travail est donc de trouver et valider une loi de comportement pour les tôles d'acier XES et d'aluminium 6016 à partir d'essais expérimentaux simples (essais de traction simple, traction plane, compression plane et cisaillement).

Dans un premier temps, nous recherchons à modéliser le comportement des matériaux étudiés (aluminium et acier); nous prenons le critère quadratique HILL 48 non centré, avec une loi d'écrouissage mixte, qui permet de tenir compte de " l'histoire du matériau " provenant de son élaboration. Les expressions théoriques des coefficients de Lankford associés $r_0$ et $r_{90}$ , ainsi que les lois gouvernant l'évolution du centre de la surface de charge, caractérisant l'écrouissage cinématique, et l'évolution de la variable d'écrouissage isotrope sont établies.

Dans un second temps, nous recherchons, à partir d'essais expérimentaux, à identifier la surface de charge décrite par le critère quadratique HILL 48 non centré. Les essais de cisaillement et de compression plane permettent d'avoir des points expérimentaux, pour l'identification de la surface de charge, dans des régions non exploitées par les essais de traction simple et plane et l'identification du comportement s'effectue alors par la caractérisation de la surface de charge (figure 1). De plus, afin d'identifier la loi d'écrouissage mixte, comportant la loi d'écrouissage cinématique et la loi d'écrouissage isotrope (Lemaître & Chaboche [4]), nous déterminons, à partir d'essais expérimentaux, les surfaces de charges pour des prédéformations uniaxiales (sens long 0°) de 8% et 14%. Ainsi, à partir des évolutions des surfaces de charges obtenues (figure 2 pour l'acier et figure 3 pour l'aluminium), nous en déduisons les coefficients caractéristiques des lois d'écrouissage mixte. Nous observons alors une bonne corrélation entre les divers essais expérimentaux et les surfaces de charges obtenues en appliquant le modèle quadratique HILL 48 non centré (figure 2 et figure 3). De même, les comparaisons expérimentales et théoriques sur un essai de traction simple (figure 4), dans le cas de l'aluminium, nous permettent de valider la loi de comportement HILL 48 quadratique non centrée avec écrouissage mixte et l'identification des divers coefficients associés.

Dans le cas de l'aluminium (figure 3), nous remarquons que la surface de charge initiale est décentrée et que cette dernière se translate progressivement et tend vers une position peu évolutive pour les grandes déformations : même si l'écrouissage cinématique reste faible par rapport à l'écrouissage isotrope, omettre ce dernier peut nous conduire à une identification approximative voir fausse du comportement des alliages d'aluminium. En revanche, le critère quadratique HILL 48 avec un écrouissage isotrope peut suffire pour identifier correctement le comportement de l'acier (figure 2).

## 1  Introduction

The lightening of vehicles is obviously one of the many goals of cars manufacturers. However, this is not the current trend, which is toward more and more complete standard equipment, optimum convenience, electronic assistance, and omnipresent strengthening structures. All these transformations contribute to the increasing weight of the vehicles. The use of aluminium in automotive/transport application is primarily driven by its high strength to weight ratio characteristics. This characteristic contributes to the efficiency in fuel consumption because of the reduction in weight. Aluminium is a serious challenge to steel : 30-40% reduction weight can be expected in using aluminium instead of steel.

The large range of available materials implies that the use of each one has to be optimised. In this way, the imposition of the metallic material of a car needs the stamping control. The stamping of thin aluminium sheets is one of the most widely used shaping processes in the car industry. The validation of stamping tools by numerical simulation needs models having realistic behaviour and reliable data concerning material. The goals are to find an "aluminium behaviour law" for numerical simulation of the stamping process and to validate the constitutive equations (plasticity and hardening criterions) based on very simple experimental tests (uniaxial tensile test, plane tensile test, plane compression test and monotonic and cyclic tensile shearing tests). The behaviour identification is carried out by characterisation of the yield surface and the determination of the elastic limit of each material.

The tensile shearing and plane compression tests give additional information for the identification of the yield surface and for both the kinematic and isotropic hardening. These experiments allowed us to obtain experimental plots in areas not studied previously, as illustrated in figure 1. Next, we can check the validity of the quadratic or non-quadratic criterion.

In this study, we are considering aluminium 6016 (Alloy 6000 series (AlMgSi). Composition in % mass: 0.9-1.5% Si., 0.4% max. Fe, 0.2% max. Cu, 0.2% max. Mn, 0.3-0.6% Mg, 0.1% max. Cr, 0.2% max. Zn, 0.15% max. Ti, Al remaining,. Heat treatment T4) and steel XES (Composition in % mass: 0.08%C max., 0.03% P max., 0.4% Mn max.).

## 2 Theoretical approach

### 2.1 Constitutive equations

Whatever the mechanical tests and the associated loading area, the later is identified in relation to a behaviour model. Such a model could be defined as follow [1]: a linear isotropic elastic behaviour (Young's modulus, Poisson's coefficient) and a plastic behaviour identified from a plastic criterion with an associated flow rule (quadratic or not) identified by the initial yield surface and an hardening model (isotropic, kinematic or both) identified by the development of the yield surface or the cyclic test. The problem is to find the constitutive equation resulting from all the realized experimental tests and corresponding to a possible physical description of all the phenomena observed during the shape up-making and the use of materials.

During the elaboration and the transformation of the metals or semi-products, the steel sheets are flattened. This flattening diminishes their thickness and gives them particular physical qualities (skin-pass process : hardening passing on rolling mills which gives an elongation of 0.5 to 2.5 %). Under the same conditions, the aluminium sheets are subject to a planage going from coils to plane sheets. These various processes concerning the shape up-making of the materials show a coupling between the plastic criterion and the hardening model (Macquaire [2]) as being a result of the "material story" (with the kinematic hardening which is very important at low strains). For metal sheets possessing orthotropy, Hill's (1948) [3] yield criterion has received the most attention and favor. This quadratic non-centered criterion imposes a mixed hardening law (kinematic and isotropic). Moreover, this criterion allows one to take into account the "material his story" resulting from its processing. So, we define the following model, called quadratic non centered Hill's (1948) yield criterion :

$$f(\sigma, X) = \sqrt{\frac{1}{2}\begin{pmatrix} H(\sigma_{11} - X_{11} - \sigma_{22} + X_{22})^2 + F(\sigma_{22} - X_{22} - \sigma_{33} + X_{33})^2 + G(\sigma_{33} - X_{33} - \sigma_{11} + X_{11})^2 + 2L(\sigma_{23} - X_{23})^2 \\ + 2M(\sigma_{31} - X_{31})^2 + 2N(\sigma_{12} - X_{12})^2 \end{pmatrix}} - R = 0 \quad (1)$$

where $\sigma$, X, R, are the strain tensor, the tensor concerning the translation of the yield locus and the isotropic hardening variable, respectively. F, G, H, L, M, N define material parameters characterizing the anisotropy.
Moreover, the mixed hardening law (kinematic and isotropic [4]) can be expressed as follow:

$$dX = C_0.d\varepsilon^p - \gamma.X.d\lambda \quad \text{and} \quad dR = C_R(R_{sat} - R).dp \quad (2)$$

where $C_0, \gamma$ are the material parameters characterizing the kinematic hardening. $C_R$, $R_{sat}$ are the material parameters characterizing the isotropic hardening. $d\varepsilon^p$, $d\lambda$ and $dp$ define the incremental strains, the constant parameter and the equivalent plastic deformation, respectively. The equivalent plastic strain can be expressed as follow (Lemaître & Chaboche [4])

$$dp = \sqrt{2\left(\frac{(F(Hd\varepsilon^p_{33} - Gd\varepsilon^p_{22}))^2 + G(Fd\varepsilon^p_{11} - Hd\varepsilon^p_{33})^2 + H(Gd\varepsilon^p_{22} - Fd\varepsilon^p_{11})^2}{(GH + FG + HF)^2} + 2\frac{d\varepsilon^{p\,2}_{12}}{N} + 2\frac{d\varepsilon^{p\,2}_{23}}{L} + 2\frac{d\varepsilon^{p\,2}_{31}}{M}\right)} \quad (3)$$

### 2.2 The mixed hardening law

We note that yield functions provide information on the properties of metals such as the orientation dependence of plastic strain ratio, the uniaxial tensile yield stress and the principal direction of strain-rate tensor. But it is not sufficient to obtain all the parameters characterizing the mixed hardening law. This is why we considered also the evolution of the Lankford coefficient, initial and subsequent yield surfaces during complex loading path, to determine all the parameters of the mixed hardening law. It was known that the directional 0° and 90° plastic strain ratios are given by $r_0 = d\varepsilon^p_2 / d\varepsilon^p_3$ and $r_{90} = d\varepsilon^p_1 / d\varepsilon^p_3$. The associated plasticity condition imposes $d\varepsilon^p_1 = \partial f/\partial\sigma_1.d\lambda$, $d\varepsilon^p_2 = \partial f/\partial\sigma_2.d\lambda$ and $d\varepsilon^p_3 = \partial f/\partial\sigma_3.d\lambda$. By substitutions, we obtain:

$$r_0 = \frac{H(X_1 - X_2 - \sigma_1) - F(X_2 - X_3)}{G(-X_3 + X_1 - \sigma_1) + F(X_2 - X_3)} \quad \text{and} \quad r_{90} = \frac{H(\sigma_2 - X_2) + 2X_1 - GX_3}{-G(X_1 - X_3) + F(\sigma_2 - X_2 - X_3)} \quad (4)$$

The objective is to obtain the four material parameters of the mixed hardening law ($C_0, \gamma, C_R, R_{sat}$). Here, we considered the evolution of the yield locus X and the isotropic hardening parameter R. To determine the evolution of the yield locus, we use the Lemaître & Chaboche model [4]:

$$dX = C_0.d\varepsilon^p - \gamma.X.d\lambda = C_0.d\varepsilon^p - \gamma.X.\psi(F,G,H,j).d\varepsilon^p_1 \quad (5)$$

where $\psi(F,G,H,j) = \sqrt{2.\frac{F(-H(j+1) - Gj)^2 + G(F + H(j+1))^2 + H(Gj - F)^2}{(GH + FG + HF)^2}}$ and $j = \frac{d\varepsilon^p_2}{d\varepsilon^p_1}$.

So, the three equations governing the yield locus evolution can be deduced:

$$X_1 = \frac{C_0}{\gamma\psi}\left(1 - e^{-\gamma\psi\varepsilon^p_1}\right) + X^0_1 e^{-\gamma\psi\varepsilon^p_1}; \quad X_2 = \frac{C_0.j}{\gamma\psi}\left(1 - e^{-\gamma\psi\varepsilon^p_1}\right) + X^0_2 e^{-\gamma\psi\varepsilon^p_1}; \quad X_3 = \frac{-C_0.(j+1)}{\gamma\psi}\left(1 - e^{-\gamma\psi\varepsilon^p_1}\right) + X^0_3 e^{-\gamma\psi\varepsilon^p_1} \quad (6)$$

We applied the same process for the determination of the evolution of the scalar R. We decide to note R as (Lemaître & Chaboche [4]) $dR = C_R(R_{sat} - R).dp$ (with $dp = \partial f/\partial(-R).d\lambda = d\lambda$). Then, we obtain $dR = C_R(R_{sat} - R).\psi(F,G,H,j).d\varepsilon^p_1$. The equation governing the evolution of the scalar R can be deduced:

$$R = R_{sat}\left(1 - e^{-C_R \psi d\varepsilon_1^p}\right) + R_0 e^{-C_R \psi d\varepsilon_1^p} \tag{7}$$

Finally, we could obtain the analytical expressions of the initial and subsequent yield surface, of the evolutions of the yield locus and the scalar R during complex determined loading path. They allow the identification of the material parameter for the obtention of the constitutive equation.

## 3 Yield surface: experimental and theoretical results

The first step is to obtain the initial yield locus. We need to identify the material parameters characterizing the anisotropy, the initial yield locus ($X_{ij}$ for i,j=1 to 3) and the scalar coefficient R. Because of the elaboration and transformation of metals (metal sheets possessing orthotropy, skin-pass…) and the definition of the Lemaître & Chaboche model, we only need four experimental tests for the identification of the initial yield locus. We decide to use the two uniaxial tensile tests (0° and 90°) and the two plane tensile tests (0° and 90°). The others tests (plane compression tests and linear and cyclic tensile shearing tests) are only used to validate definitively the initial yield surface. By considering the quadratic non centered Hill's yield criterion (1), the associated plasticity condition and the specific conditions for each tensile test (uniaxial 0° tensile test: $\sigma_2 = \sigma_3 = 0$; uniaxial 90° tensile test: $\sigma_1 = \sigma_3 = 0$; plane strain 0° tensile test: $d\varepsilon_2^p = 0$ and $\sigma_3 = 0$; plane strain 90° tensile test: $d\varepsilon_1^p = 0$ and $\sigma_3 = 0$), we obtain the expressions:

$$\sigma_1^{uniaxial} = X_1 + \frac{-(GX_3 + HX_2) + \sqrt{(GX_3 + HX_2)^2 - 2\left[(F+H)X_2^2 + (G+F)X_3^2 - 2FX_2X_3 - 2R^2\right]}}{2} \tag{8}$$

$$\sigma_2^{uniaxial} = X_2 + \frac{-(HX_1 + FX_3) + \sqrt{(HX_1 + FX_3)^2 - (F+H)\left[2X_1^2 + (G+F)X_3^2 - 2GX_1X_3 - 2R^2\right]}}{F+H} \tag{9}$$

$$\sigma_1^{plane} = X_1 + \frac{-X_3(G + HF/(F+G)) + \sqrt{X_3(G + HF/(F+G))^2 - (2 - H^2/(F+H))\left[X_3^2(G + F - F^2/(F+H)) - 2R^2\right]}}{(2 - H^2/(F+H))} \tag{10}$$

$$\sigma_2^{plane} = X_2 + \frac{-X_3(HG + 2F) + \sqrt{X_3^2(GH + 2F)^2 - 4(-H^2/2 + F + H)\left[X_3^2(G + F - G^2/2) - 2R^2\right]}}{2(F + H - H^2/2)} \tag{11}$$

By using equations (4), (8), (9), (10) and (11), we obtain the expression of the quadratic non centered Hill's (1948) yield surface. Figure 2 and Figure 3 present the different results obtained for steel and aluminium, respectively. One can observe a perfect correlation between the experimental tests and the yield surface. Hence, we note that the yield locus for aluminium does not remain centered, as illustrated in Figure 3 : this implies the need of kinematic law.

In order to obtain the constitutive law of steel and aluminium, it is necessary to have subsequent experimental yield surfaces during complex loading. We carry out predeformations on steels (8% and 14% uniaxial deformations 0°), which were then cut to perform classical experimental tests for the yield surface (uniaxial tensile tests, plane tensile test, etc...). We decide to use two uniaxial tensile tests (0° and 90°), two plane strain tensile tests (0° and 90°) to determine the subsequent yield surface. The other data (plane compression tests, a cyclic tensile shearing test and the evolutions of the Lankford coefficients) are only used to validate definitively the yield surfaces. Figure 2 and Figure 3 present the results obtained for steel and aluminium, respectively. One can observe a perfect correlation between all the experimental tests and the yield surface. Hence, we note that the evolution of the yield surface during the complex imposed loading path has a predominant isotropic form. However, in the case of aluminium, we observe that the initial yield center is different from zero and we have a evolution of this yield center for the subsequent yield surfaces.

## 4 Determination of the mixed hardening law

In this part, we consider the mixed nonlinear hardening law (Lemaître & Chaboche) defined previously. We intend to identify the four coefficients of this law ( $C_0, \gamma, C_R, R_{sat}$). We previously determined the initial yield surfaces and the evolution of the yield surfaces during a complex loading path. Using the evolution of the yield surfaces, it is easier to determine the evolution of the yield locus and the evolution of the scalar R. Therefore, we could use a lot of tests to identify the four coefficients and to validate them definitively. At the beginning, we introduce the evolution of the yield locus and the evolution of the scalar R. We obtain also the four coefficients of the mixed non-linear hardening law "Lemaître & Chaboche non-linear". The values of the coefficients for aluminium and steel are given in Table 1. Secondly, we have a good agreement between the experimental uniaxial 0° tensile test and the theoretical expression associated, as illustrated in figure 4.

## 5 Conclusion

In this study, the "Hill 48 quadratic yield criterion with both isotropic and kinematic hardening ". By comparing both experimentally measured and calculated data based on this criterion, it is demonstrated that this criterion leads to a good description of the phenomena. The use of the characterisation of yield surface is necessary to have a good representation of the constitutive equation because of the sensitivity of models to plastic strain ratios. In this paper, we explain that the determination of the elastic limit of each material and the use of all mechanical tests (uniaxial tensile test, plane tensile test, plane compression test, linear and cyclic tensile shearing tests) are used to confirm the "Hill 48 quadratic yield criterion with both isotropic and kinematic hardening" behaviour law.

|  | Anisotropic coefficients | | | | Lankford coefficients | | |
|---|---|---|---|---|---|---|---|
|  | F | G | H | N | $r_0$ | $r_{45}$ | $r_{90}$ |
| Steel XES | 0.75 | 0.8 | 1.2 | 2.7 | 1.77 | 1.35 | 1.98 |
| Aluminium 6016 | 1.3 | 1.15 | 0.85 | 2.6 | 0.68 | 0.41 | 0.65 |

|  | Kinematic hardening variables | | | | | Isotropic hardening variables | | |
|---|---|---|---|---|---|---|---|---|
|  | $X_1^0$ | $X_2^0$ | $X_3^0$ | $C_0$ | $\gamma$ | $R_0$ | $C_R$ | $R_{sat}$ |
| Steel XES | 9 | 0 | -9 | 30 | 1 | 145 | 26 | 360 |
| Aluminium 6016 | 3 | -1.4 | -1.6 | 240 | 13.5 | 106 | 11 | 250 |

Table 1: Coefficients for the mixed hardening law
Table 1: Coefficients de la loi d'écrouissage mixte

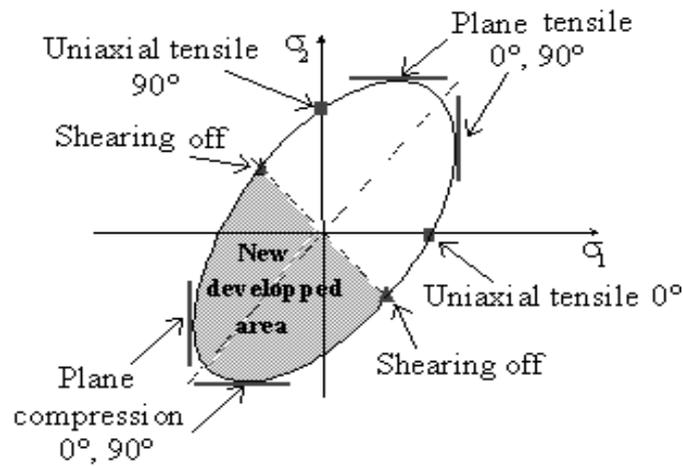

Figure 1 : Localization of the experimental tests
Figure 1 : Localisation des essais expérimentaux

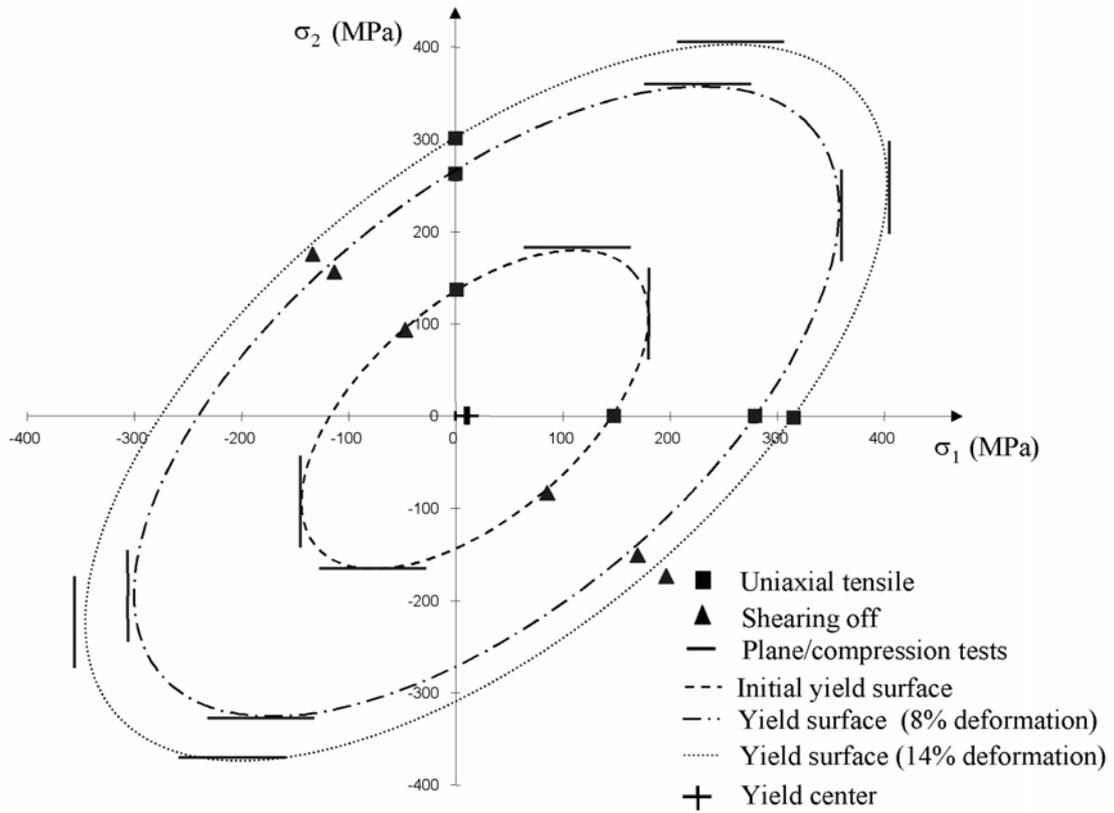

Figure 2 : Evolution of yield surface (steel XES)

Figure 2 : Evolution de la surface de charge (acier XES)

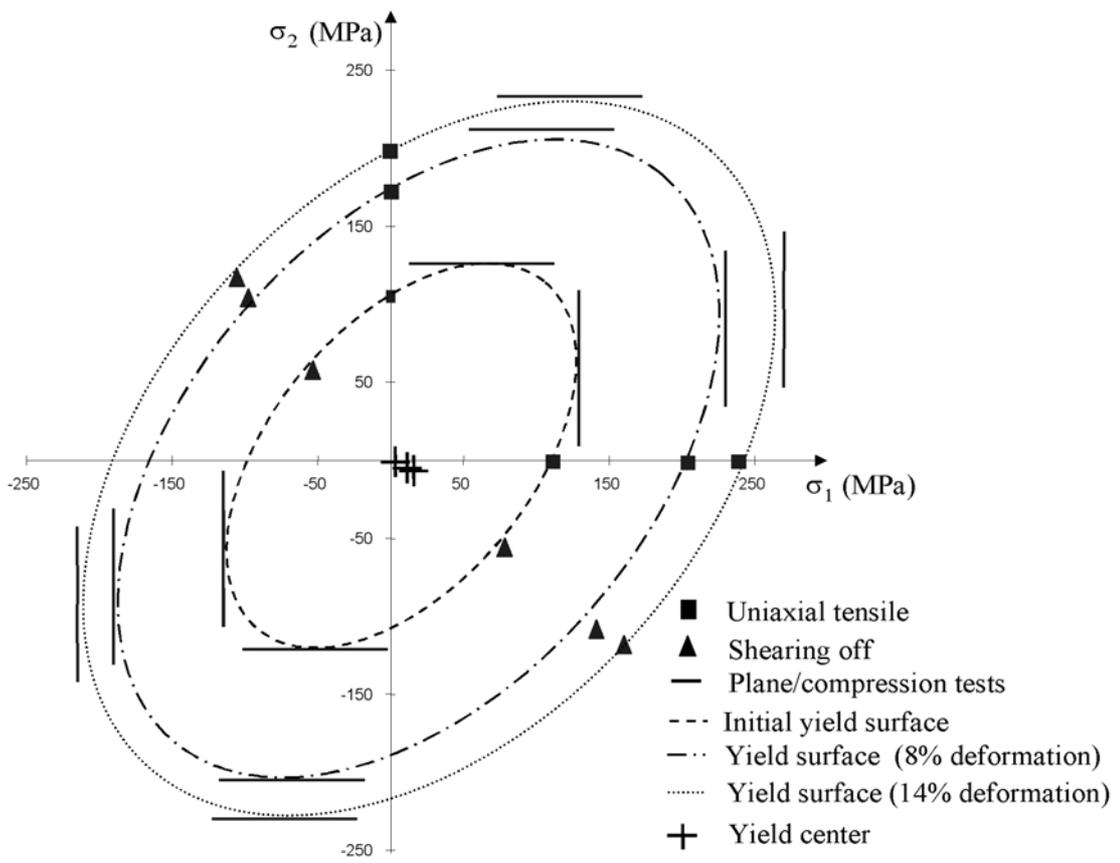

Figure 3 : Evolution of yield surface (aluminium 6016)
Figure 3 : Evolution de la surface de charge (aluminium 6016)

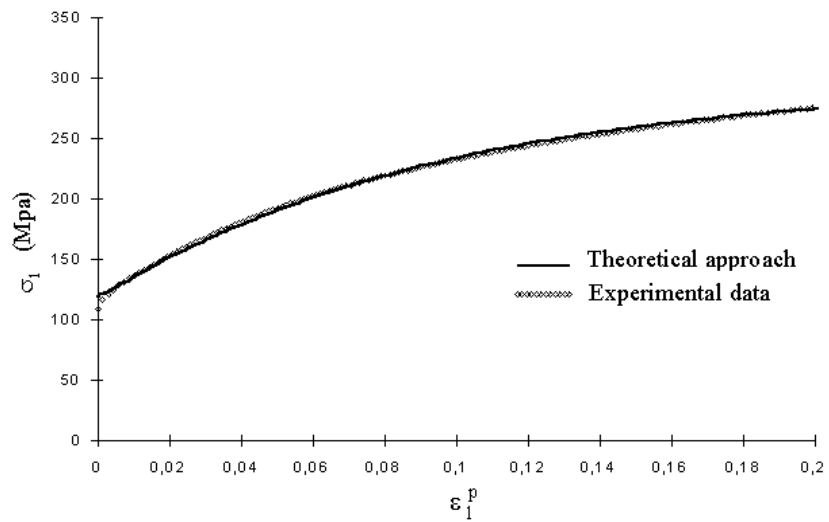

Figure 4 : Correlation with an uniaxial tensile test 0° (Aluminium 6016)
Figure 4 : Corrélation avec un essai de traction simple 0° (Aluminium 6016)